\documentclass[a4paper,aps,11pt,preprint]{revtex4}
\usepackage{graphicx,times,nicefrac,amsmath,amsfonts,amssymb,amsthm,epsf,bm,bbm,longtable,dcolumn}
\usepackage{revsymb,wasysym,mathrsfs,color}
\DeclareGraphicsExtensions{.pdf}
\begin{document}

\title{Main magnetic focus ion source with the radial extraction of ions}

\author{V. P. Ovsyannikov}\thanks{URL: \url{http://mamfis.net/ovsyannikov.html}}
\affiliation{Hochschulstr. 13, D-01069  Dresden, Germany}

\author{A. V. Nefiodov}\thanks{E-mail: anef@thd.pnpi.spb.ru}
\affiliation{Petersburg Nuclear Physics Institute, 188300 Gatchina, St.~Petersburg, Russia}

\widetext

\begin{abstract}
In the main magnetic focus ion source, atomic ions are produced in the local ion trap created by the rippled electron beam in focusing magnetic field. Here we present the novel modification of the room-temperature hand-size device, which allows the extraction of  ions in the radial direction perpendicular to the electron beam across the magnetic field. The detected X-ray emission evidences the production of  Ir$^{44+}$ and Ar$^{16+}$  ions. The ion source can operate as the ion trap for X-ray spectroscopy, as the ion source for the production of highly charged ions and also as the ion source of high brightness.
\end{abstract}
\maketitle

\section{Introduction}

In this work, we present the miniature main magnetic focus ion source (MaMFIS) with the radial extraction of highly charged ions. The design of this device is based on fundamental dependence of the potential distribution of axially symmetric electron beam, which is determined by logarithmic ratio of the radius $R$ of the drift tube  to the radius  $r_e$ of the electron beam. The sag of the potential $\Delta U$ in space between the metal wall of the drift tube and the $z$ axis of the electron beam reads 
\begin{equation} \label{eq1}
\Delta U(r_e) = V_e \left(1+2 \ln \frac{R}{r_e}\right),
\end{equation}
where the potential difference inside the electron beam $V_e$ is given by
\begin{equation} \label{eq2}
V_e=  \frac{U P}{4 \pi \varepsilon_0 \sqrt{2 \eta}} .
\end{equation}
Here $\varepsilon_0$ is the permittivity of free space, $\eta = e/m$ is the magnitude of the electron charge-to-mass ratio, $U$ is the potential of the drift tube with respect to the potential of the cathode, $P= I_{e}/U^{3/2}$ is the perveance of the electron beam and $I_{e}$ is the electron current. 

The local ion traps, which are formed in crossovers of the rippled electron beam with the variable radius $r_e(z)$, are used for the efficient production and axial extraction of highly charged ions in the MaMFIS \cite{1,2,3}. The local ion traps are characterized by extremely high electron current density and relatively short length of the trap (typically of about 1 mm). It implies that ions in the trap are concentrated in a very small volume and can be extracted in the radial direction perpendicular to the electron beam through the extractor electrode with small entrance diameter. This concept allows one to create the very compact ion source of highly charged ions with high brightness, especially for elements of the first half of the periodic table.

\begin{figure}[tbhp]
\centering
\includegraphics[width=0.62\columnwidth,trim=30  60  40  310,clip]{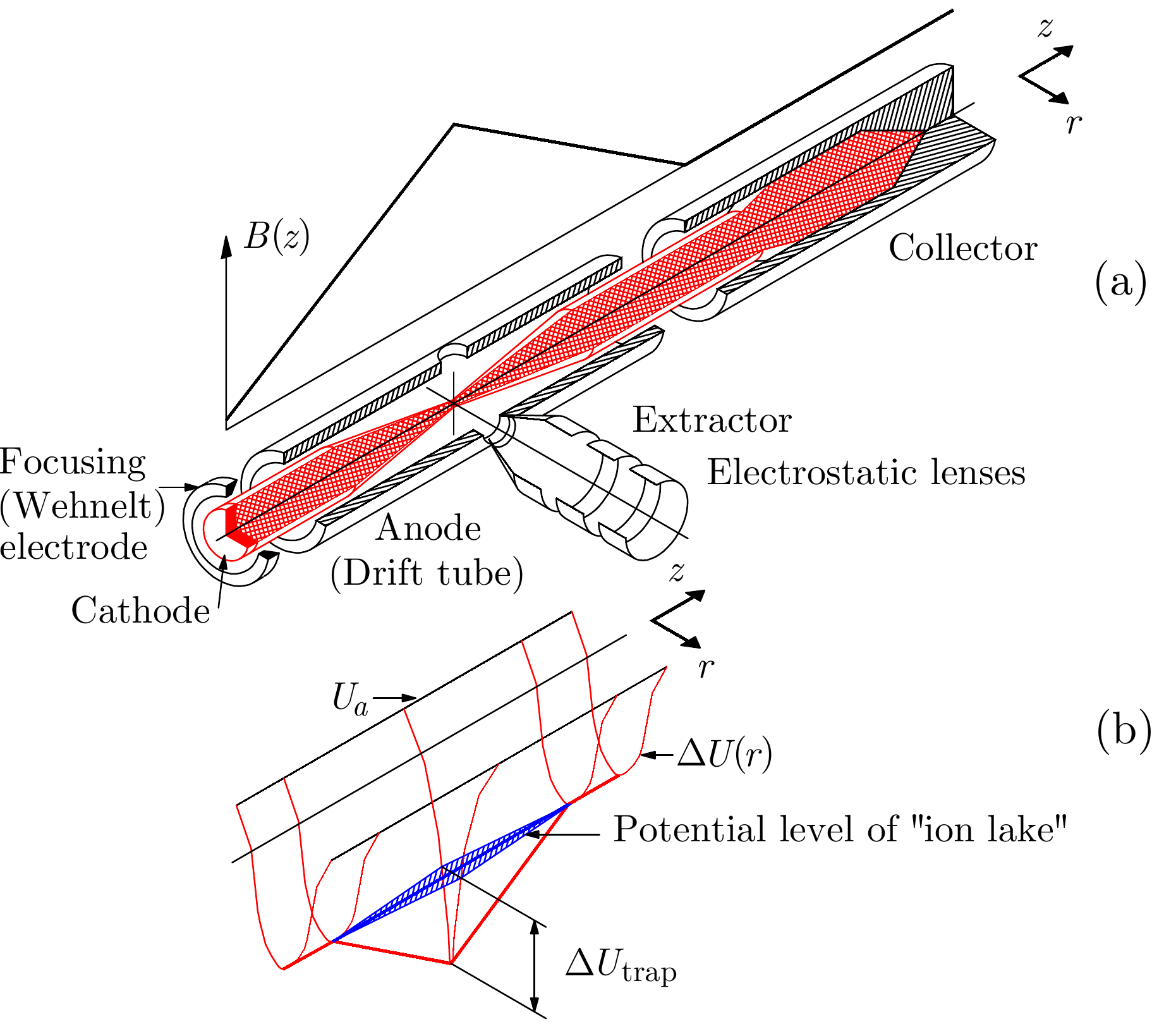}
\caption{\label{fig1} (Color online) The principal scheme of the ion source. The $z$ axis indicates the direction of electron beam, $B(z)$ is the distribution of focusing magnetic field, $U_a$ is the potential of anode (drift tube),  $\Delta U(r)$ is the sag of potential in the space of drift tube  and $\Delta U_\mathrm{trap}$ is the depth of local ion trap.}
\end{figure}

\section{Principle of design}

The principle of design of the MaMFIS with the radial extraction of ions is shown in Fig.~\ref{fig1}(a).  The ion source includes the electron gun, the anode integrated with the drift tube, the electron collector and the extractor with ion optics.  The anode has few holes in the middle plane.  The extractor is located in one of these holes and is installed in the direction perpendicular to the electron beam.

\begin{figure}[t]
\centering
\includegraphics[width=0.64\columnwidth,trim=30  60  30  360,clip]{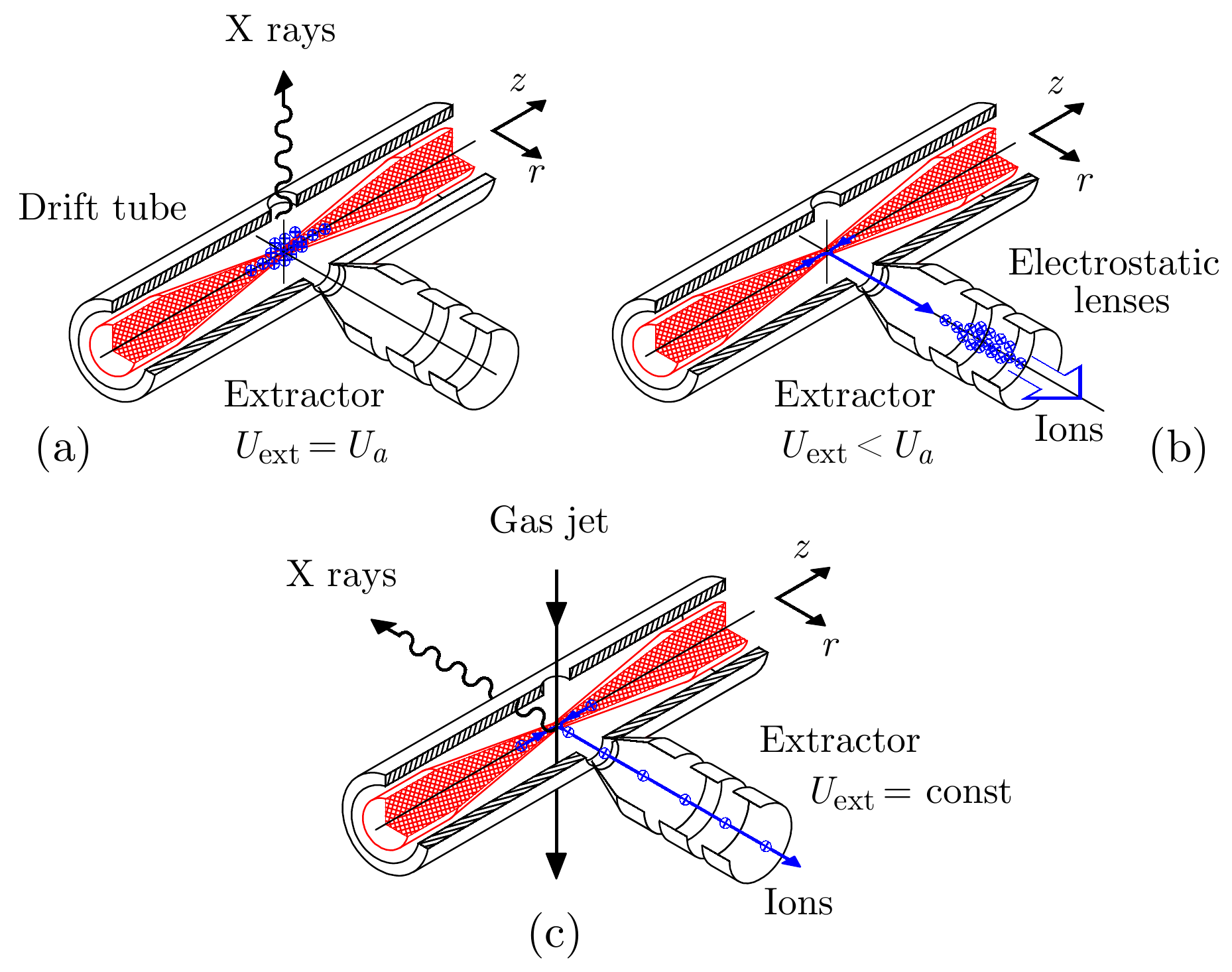}
\caption{\label{fig2} (Color online)  Running modes:  ion trap (a); ion source (b); ion source of high brightness (c).}
\end{figure}

The beam of electrons emitted from cathode is focused by magnetic field $B$ and Wehnelt electrode into the sharp focus in the median plane of anode. The potential distribution on the length of anode is shown in  Fig.~\ref{fig1}(b).  In three-dimensional space, the potential surface along the $z$ axis looks like a curved gutter with the greatest depth at the point corresponding to the position of crossover of the electron beam characterized by the smallest radius $r_\mathrm{min}$. The potential well, which serves as a trap for storage of atomic ions, occurs due to variability of the radius $r_e$ of the electron beam along the $z$ axis within the range $r_\mathrm{min} \leqslant r_e \leqslant r_\mathrm{max}$. Ions can fill the ion trap up to the potential level corresponding to the electron beam with the constant  maximal  radius $r_\mathrm{max}$. This potential level restricts the depth of the ``ion lake'', which is estimated by \cite{2}
\begin{equation} \label{eq3}
\Delta U_\mathrm{trap} = 2 V_e \ln \frac{r_\mathrm{max}}{r_\mathrm{min}} .
\end{equation}
The ions with energies above this crucial level leave the trap in the axial direction.

\begin{figure}[t]
\centering
\includegraphics[width=0.64\columnwidth,trim=30  60  30  520,clip]{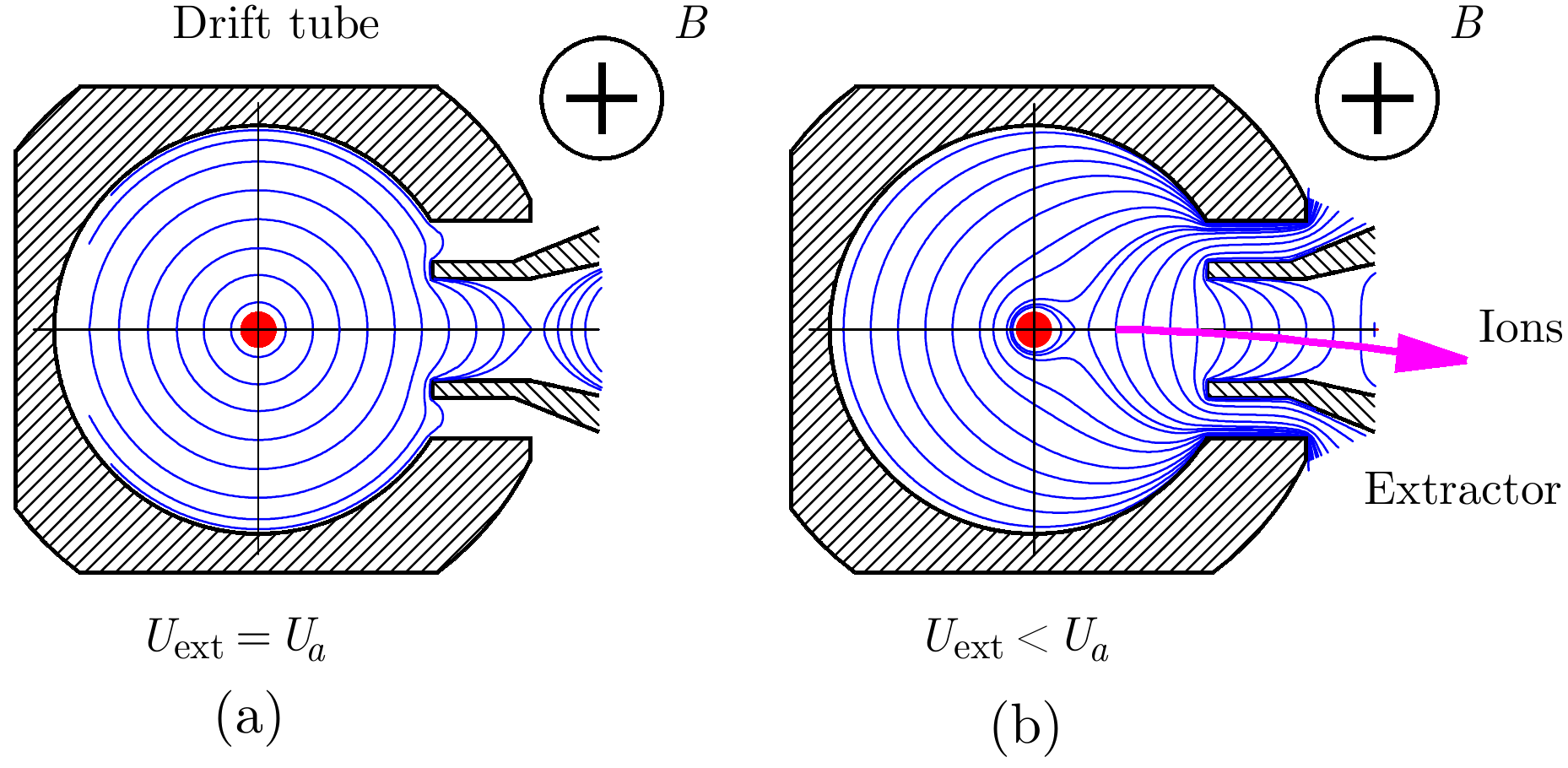}
\caption{\label{fig3} (Color online) Equipotential lines of electric field in the cross section of drift tube. 
Trapping mode (a); extraction mode (b).}
\end{figure}

\begin{figure}[tbh]
\includegraphics[height=6.5cm,keepaspectratio]{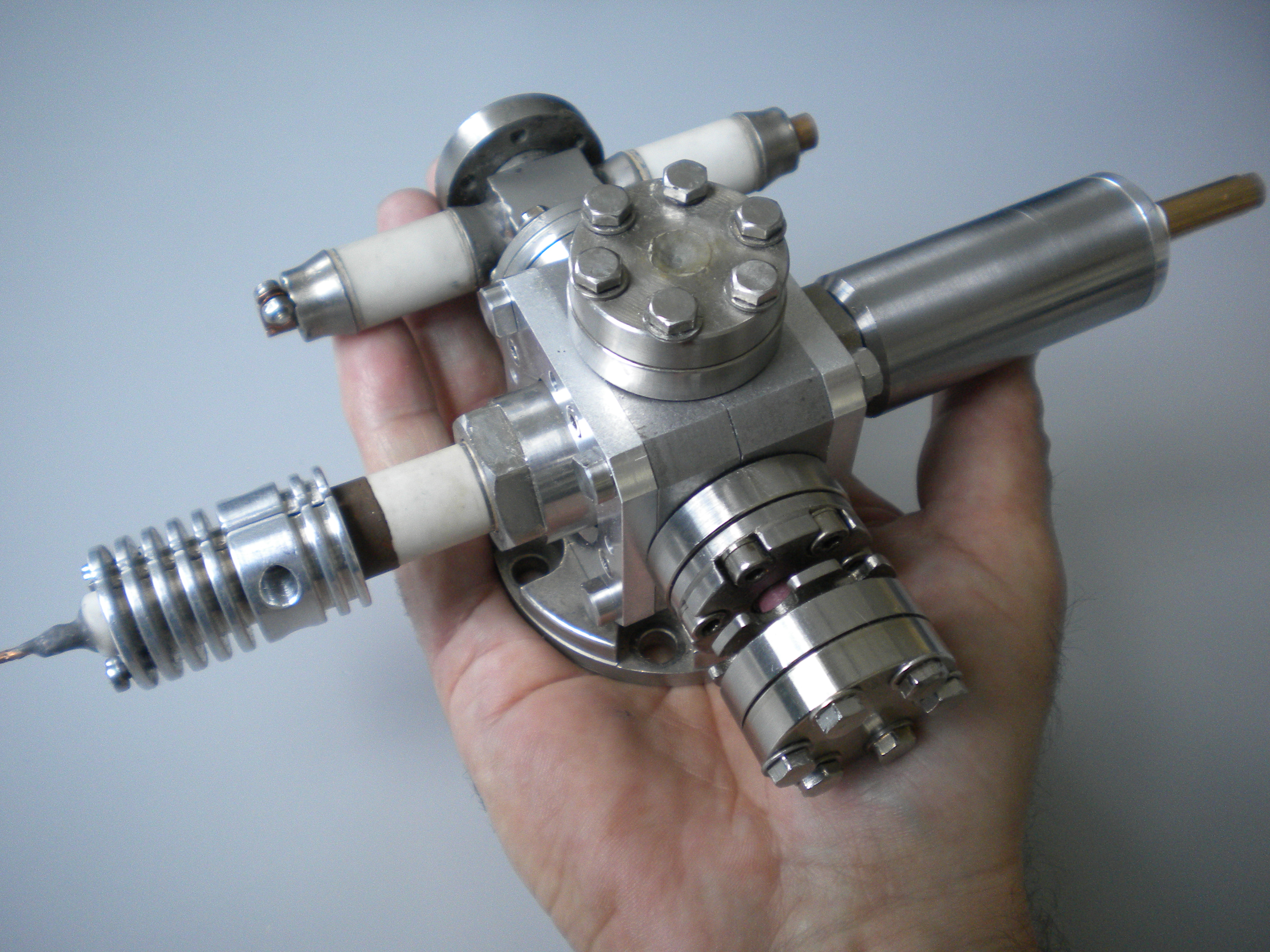}
\hfill
\includegraphics[height=6.5cm,keepaspectratio]{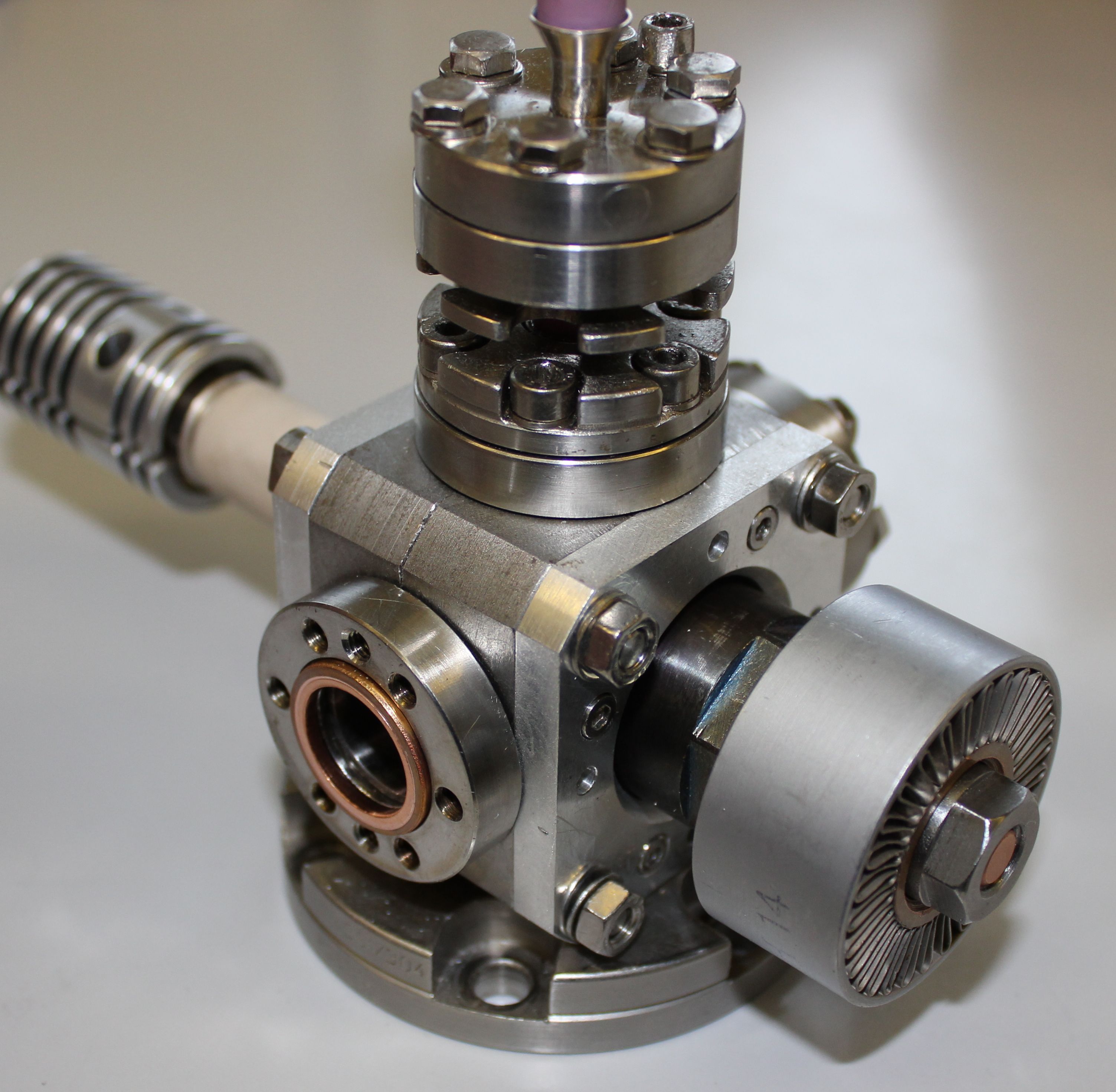}
\caption{\label{fig4} (Color online) General view of the MaMFIS with the radial extraction of ions. Device with water cooling collector for dissipation power of up to 1 kW (left photo) and with air cooling collector for dissipation power of up to 300 W (right photo).}
\end{figure}

The installation can operate in three different running modes. The mode of ion trap is realized, if the potential $U_\mathrm{ext}$ of the extractor is equal to the potential  $U_{a}$ of the drift tube  [see  Fig.~\ref{fig2}(a)]. This running mode can be used for X-ray spectroscopy and microplasma research. The running mode of ion source is switched on, when the potential of the extractor is lower than the potential of the drift tube  [see  Figs.~\ref{fig2}(b) and  \ref{fig2}(c)]. In this case,  ions escape the trap in the radial direction across the magnetic field. The ion extraction of this type has been thoroughly studied in ion sources for cyclotrons. The extraction potential can be either constant or pulsating. The pulsating potential corresponds to the running mode of ion traps with a certain confinement time. The repetition rate, i.e. the time between pulses of the extraction voltage, determines the confinement time, while duration of the voltage pulse determines the duration of the ion pulse. The running mode with the constant extraction potential allows the production of the direct ion current from extremely small ionization volume. The relatively intense ion current is gained under poor vacuum in the ionization region. Gas jet can be applied for this purpose. Some deviation of ions from the extraction axis due to influence of the magnetic field is compensated by the electrostatic lenses (see  Fig.~\ref{fig3}).

\section{Design of pilot example}
 
The pilot example of the miniature ion source was developed in two modifications: for relatively high voltage (up  to 10 kV) with the water cooling collector and for low voltage (up to 2 kV) with the air cooling (see photos in Fig.~\ref{fig4}). The ion source consists of the vacuum body, the electron gun, the magnetic focusing system, the electron collector and the extraction system. The vacuum vessel is the double cross of the minimum standard 16-mm-ConFlat flanges.

\begin{figure}[tbhp]
\centering
\includegraphics[width=0.5\columnwidth,trim=40  50 50  60,clip]{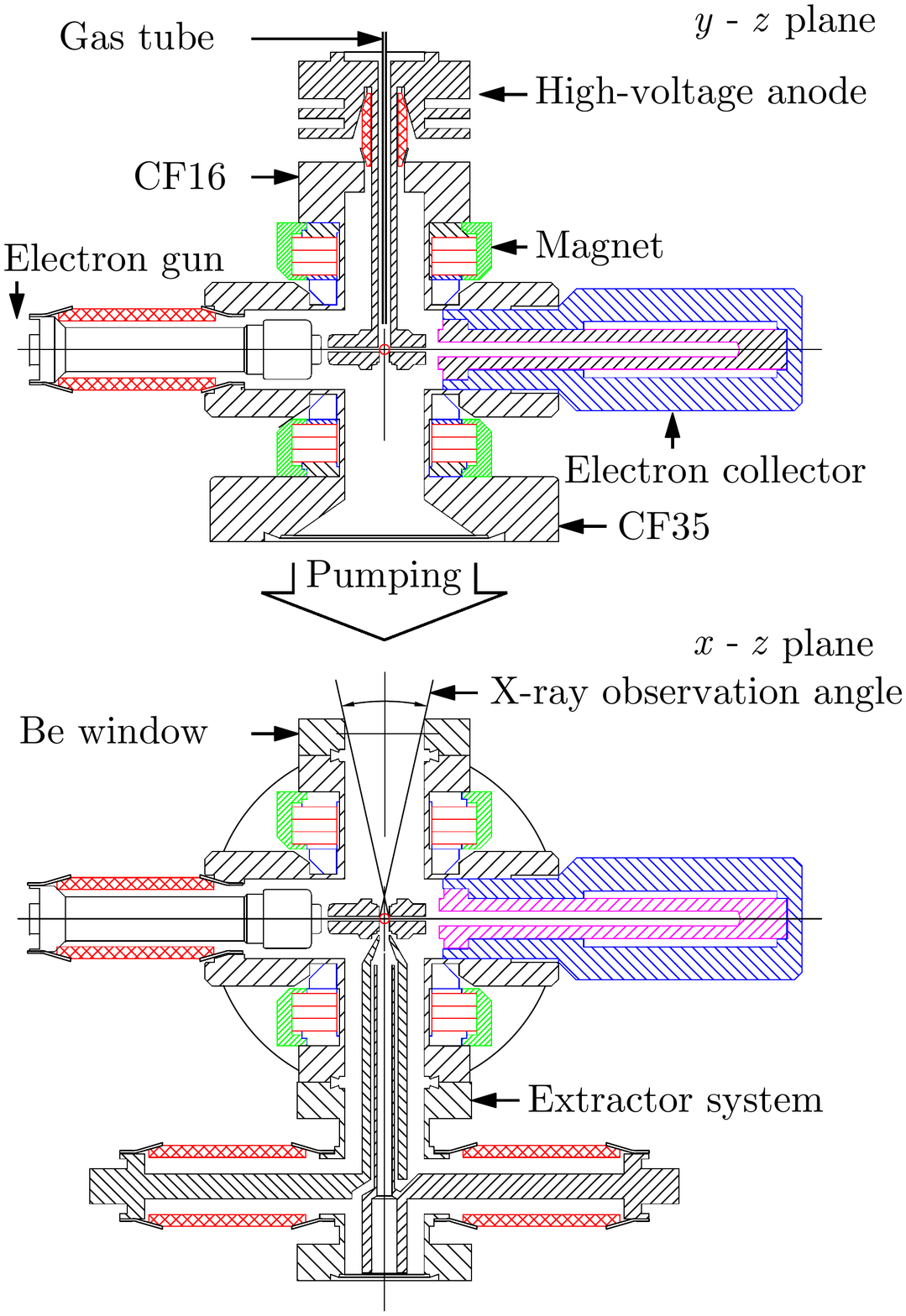}
\caption{\label{fig5} (Color online)  Sketch of the ion source.}
\end{figure}

\begin{figure}[tbhp]
\includegraphics[height=0.4\columnwidth,angle=0,keepaspectratio]{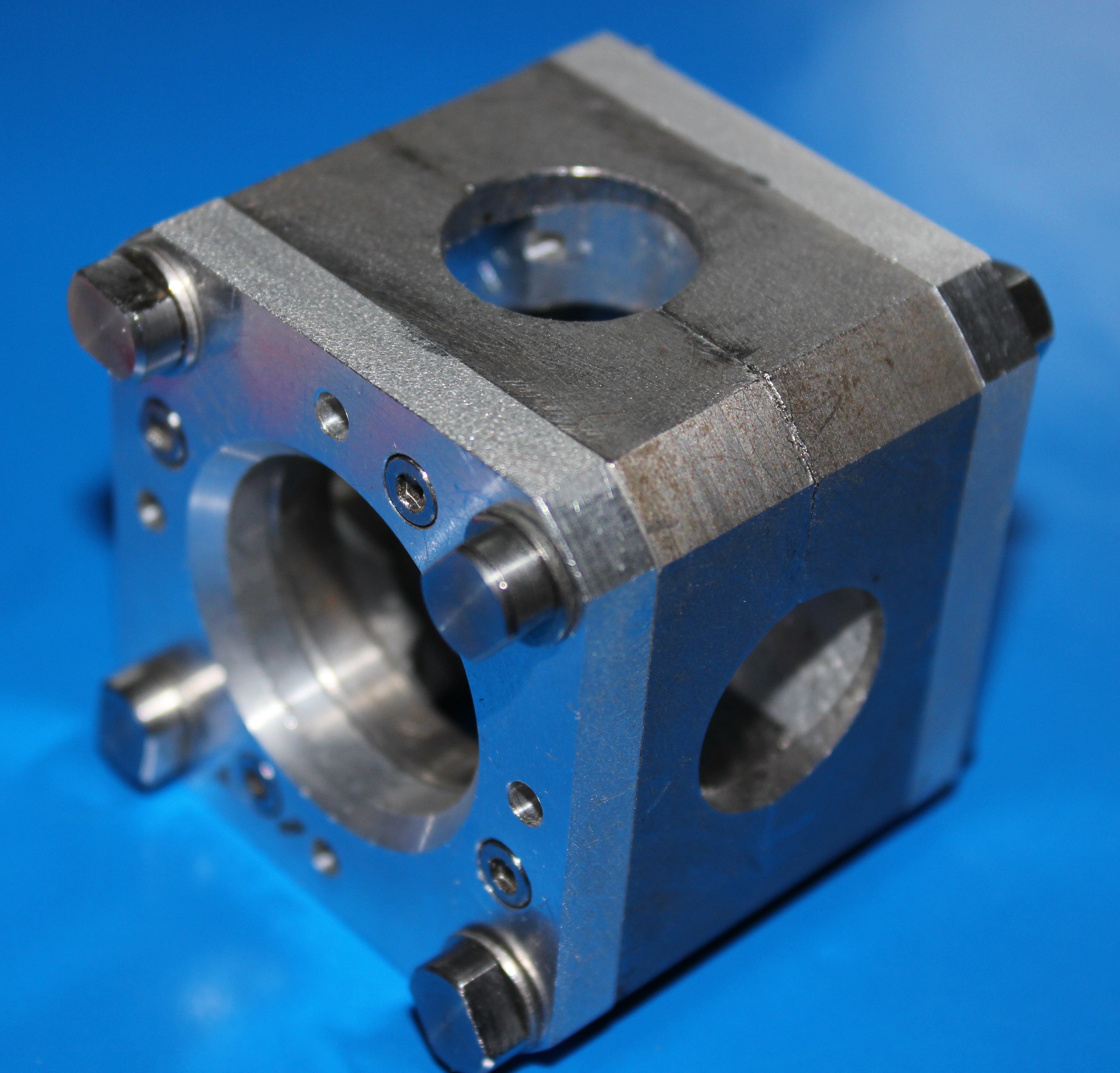}
\hfill
\includegraphics[width=0.57\columnwidth,angle=0,keepaspectratio,trim=20  20  10  0,clip]{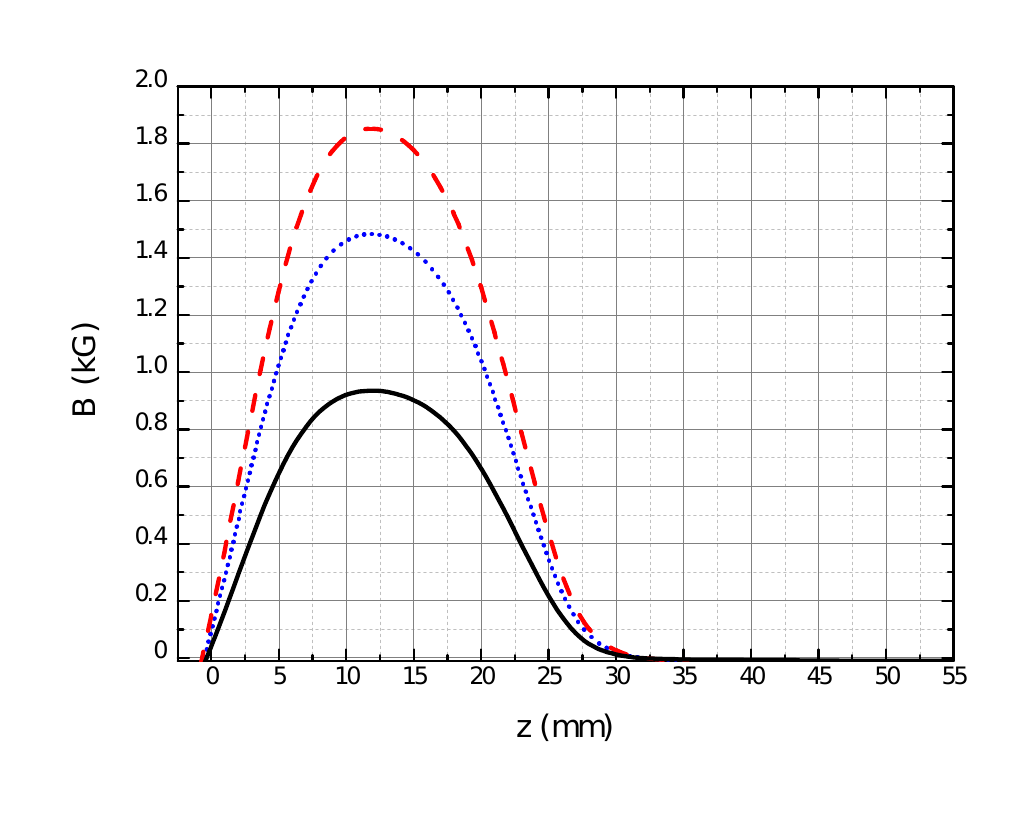}
\caption{\label{fig6} (Color online)  Permanent magnet focusing system. General view (photo) and possible distributions of the magnetic field along the $z$ axis of the ion source (graphs).}
\end{figure}

The principle sketch of the ion source is shown in Fig.~\ref{fig5}. The cross sections of the construction are made in the Cartesian coordinate system. The electron gun and the electron collector are installed in the $y$-$z$ plane. The high-voltage anode integrated with the drift tube is located between the cathode and the collector. The drift tube has three holes in the central part. The extractor electrode with the ion optics is mounted in the direction perpendicular to the drift tube. The Be window for X-ray spectroscopy is placed opposite to the radial extractor.

The magnetic focusing system can be considered as a key part of the device, since the magnitude of the magnetic field and its distribution along the $z$ axis define properties of the electron beam. Here it is used the principle of focusing systems with the radial permanent magnets (see, for example, Ref.~\cite{4}). The magnetic field system consists of two iron parts of complicated shape, in which the rectangular permanent magnets are held [see photo in Fig.~\ref{fig6}]. These parts are fastened to each other with screws on the vacuum body of the ion source. The radial magnetization vectors of the permanent magnets in each part have opposite directions. The number of permanent magnets and their positioning in the system can be changed in various combinations. Therefore, both the amplitude value and distribution of the magnetic field can be discretely changed [see graphs in Fig.~\ref{fig6}].

\section{Electron beam}

\begin{figure}[t]
\centering
\includegraphics[width=0.75\columnwidth,trim=0  40  0  445,clip]{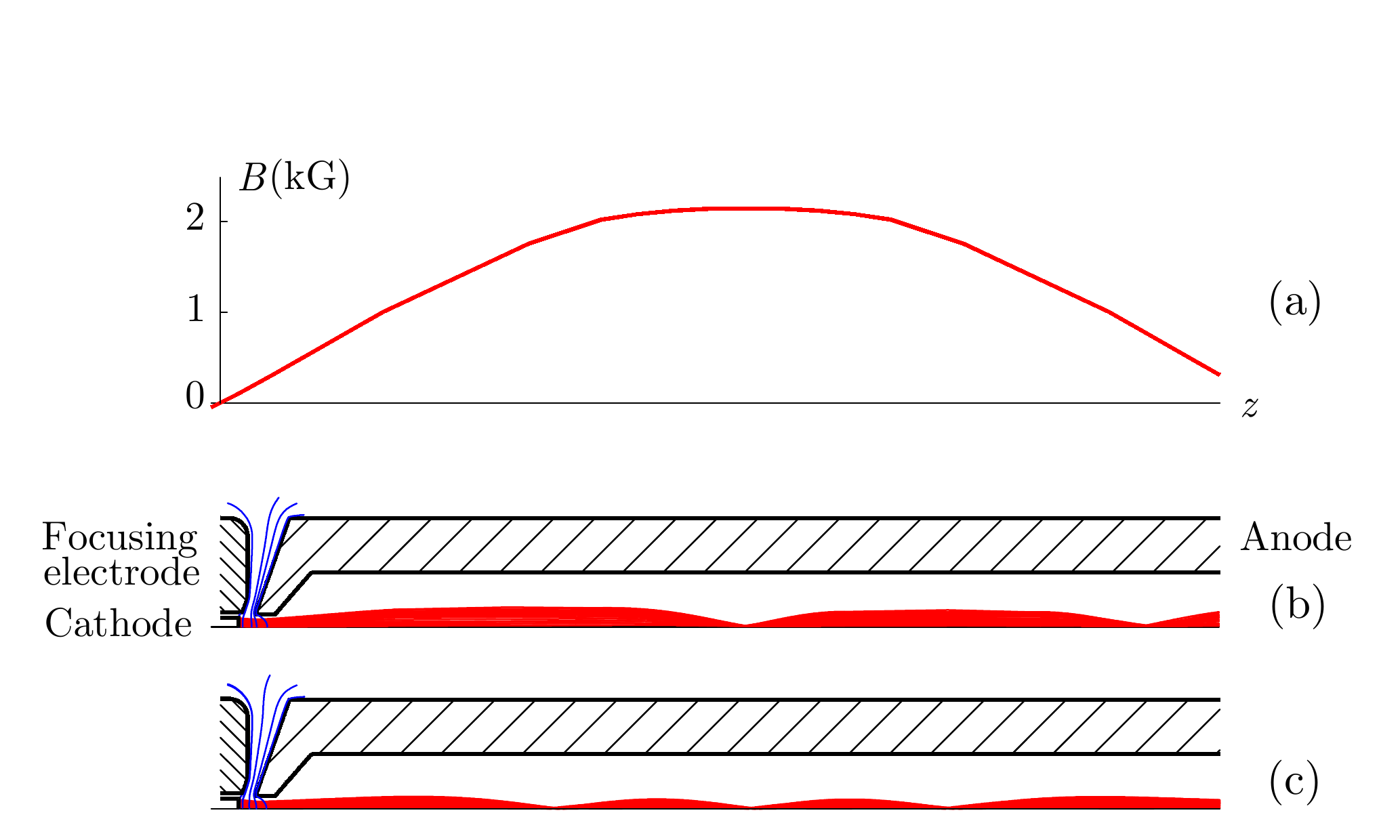}
\caption{\label{fig7} (Color online)  Magnetic field distribution (a) and electron trajectories for the following parameters: $I_e= 50$~mA, $E_e=4$~keV (b) and $I_e=25$~mA, $E_e=2.5$~keV (c).}
\end{figure}

The electron beam is created in the electron gun with the metal alloy Ir-Ce ca\-tho\-de of 0.5 mm in diameter. Then it is refracted by magnetic focusing system into a long flow, which can have different properties. According to the basic requirements the electron beam with one focus in the drift tube is suitable for the ion source. However, for typical sizes of the installation with the vacuum standard CF16 in the electron energy range of about 1 keV the focusing magnetic field should be either very short or very weak. Therefore, here we shall consider the electron beam with a few successive focuses in a thick magnetic lens. The trajectories of the electron beam calculated for different parameters are shown in Fig.~\ref{fig7}  \cite{5}. In order to run the MaMFIS in the first focus of the electron beam with the energy of 4 keV,  the magnetic field distribution should correspond to the curves in Fig.~\ref{fig6}. According to theoretical estimates, the electron current density lies within the range of  5--10~kA/cm$^2$  for high voltage  and is about  50~A/cm$^2$ for low electron energies.

\begin{figure}[t]
\centering
\includegraphics[width=0.65\columnwidth,trim=20  10  10  20,clip]{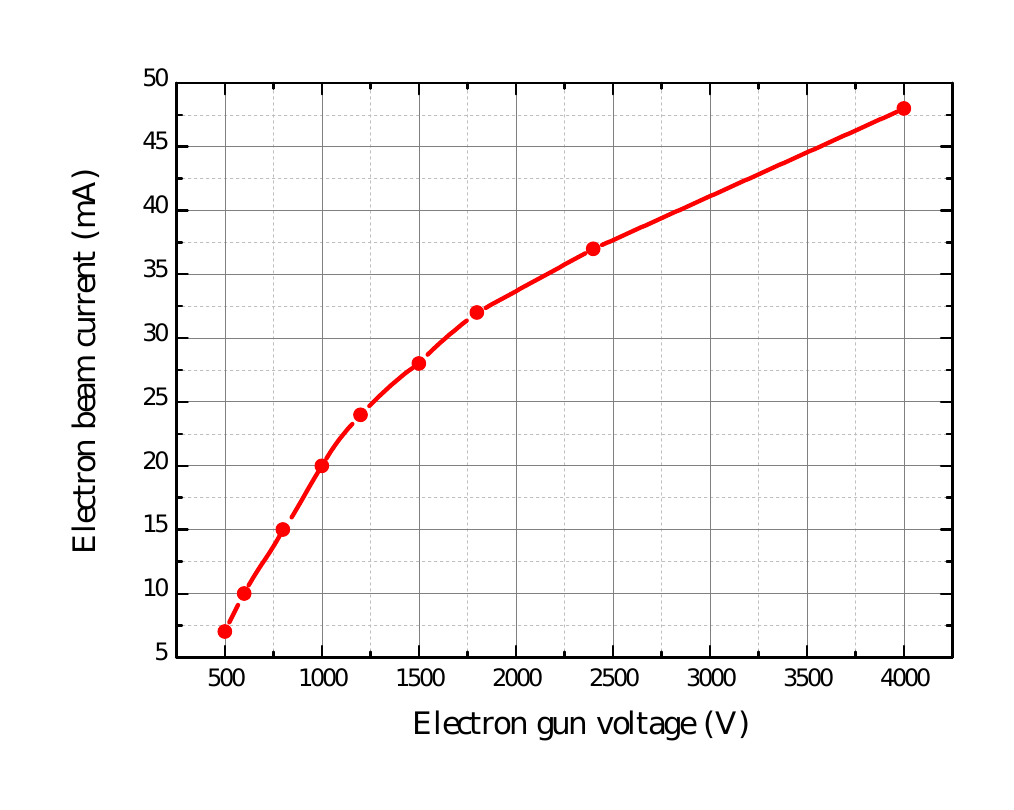}
\caption{\label{fig8} (Color online) Experimental volt-ampere characteristics of the electron gun.}
\end{figure}

One important conclusion follows from the performed computations. The proposed method of ionization is implemented for the particular values of  energy of the electron beam. Changing distributions of the magnetic field can extend the number of operating points. This possibility was taken into account in design of the magnetic focusing system. In the first experiments,  the current $I_e$ and the energy $E_e$ of the electron beam were smoothly changed within the ranges of 10-50~mA and 0.6-4~keV, respectively, without significant interceptions of the current (see Fig.~\ref{fig8}).

\begin{figure}[tbh]
\centering
\includegraphics[width=0.65\columnwidth,trim=20  10  10  15,clip]{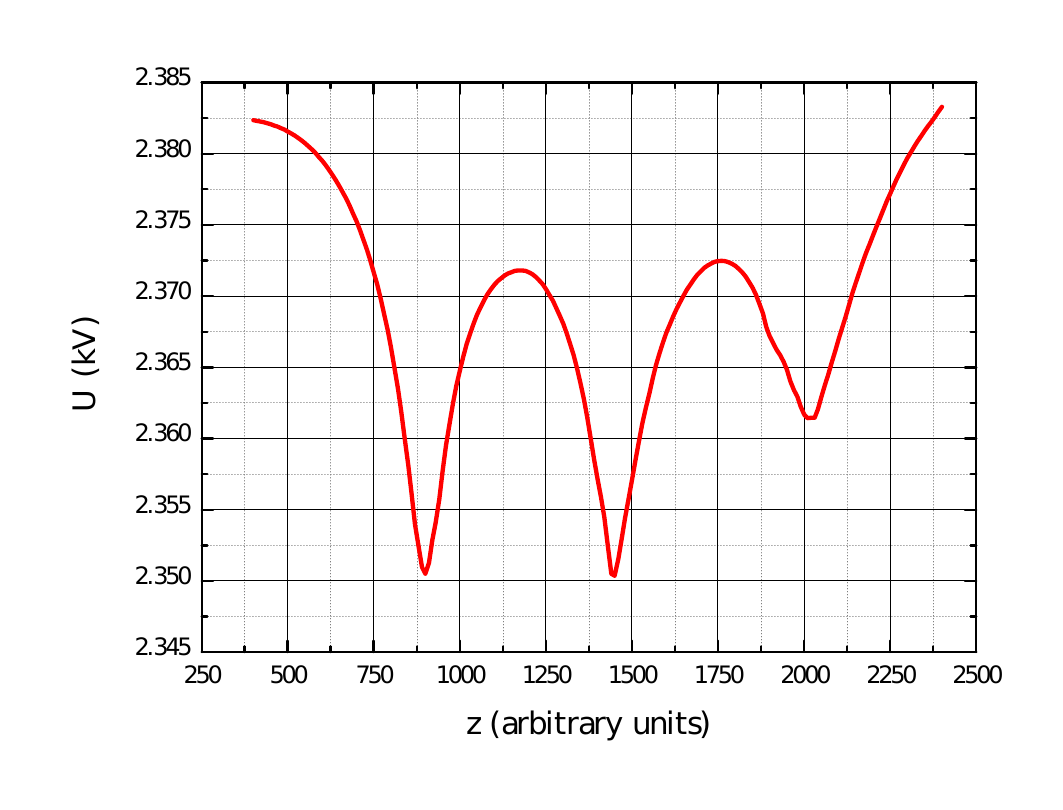}
\caption{\label{fig9} (Color online) Potential distribution on the $z$ axis of rippled electron beam characterized by   
$I_e= 25$~mA and $E_e=2.5$~keV.}
\end{figure}

\section{Local ion traps}

As we mentioned above, the rippled electron beam propagating along the cylindrical drift tube creates a sequence of local ion traps, the depths of which are characterized by the ratio $r_\mathrm{max}/r_\mathrm{min}$  of radii of the electron beam. The positions of the local ion traps depend on the energy of electron beam and the focusing magnetic field. Since the characteristics of the local ion traps are difficult to measure experimentally,  we have made theoretical prediction of the potential distribution on the $z$ axis of  electron beam.

In Fig.~\ref{fig9}, it  is calculated the potential distribution, which corresponds to the electron beam with three focuses [see Fig.~\ref{fig7}(c)]. The distribution has two sharp focuses, while the third focus is relatively vague. If the ion extractor is located in front of the second focus, the emission area for ions turns out to be very small. Accordingly, the ion source is characterized by high brightness. It should be noted that ions from the second focus need to be extracted only. The ions from the first and third ion traps overflow into the area of extraction, because the depths of these traps are significantly reduced in the process of ion compensation. The figure illustrates also the main problem of multiple-focus ion traps. The radius of electron beam at the point of the third focus becames larger than those in previous focuses. This feature has been studied by K. Ambos \cite{6}. The rippled electron beam is smoothed rapidly on some distance from the electron gun under influence of aberrations of anode lens, thermal velocities and other effects. Therefore, design of the MaMFIS with the rippled electron beam with more than three focuses is inefficient.

\section{X-rays emission by ions of cathode materials}
\subsection{Electron beam with the energy of 4 keV}

\begin{figure}[tbhp]
\begin{minipage}[t]{\textwidth}
\centering\includegraphics[width=0.65\columnwidth,angle=0,keepaspectratio,trim=20  10  10  10,clip]{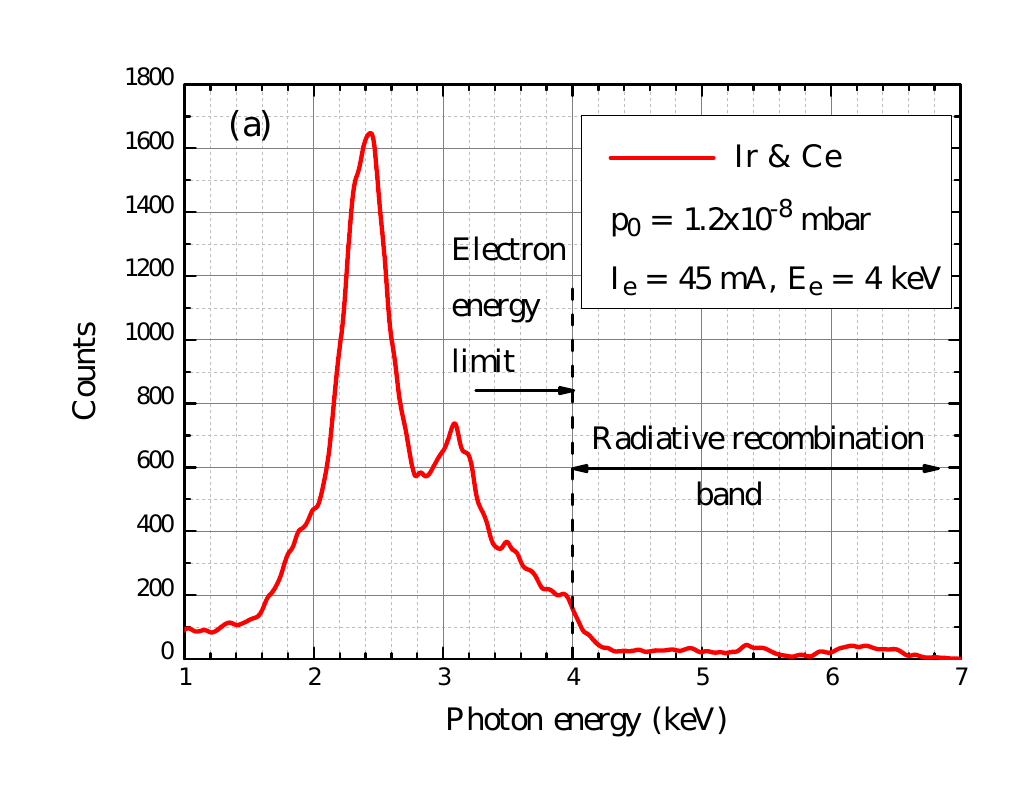}
\end{minipage}
\begin{minipage}[b]{\textwidth}
\centering\includegraphics[width=0.65\columnwidth,angle=0,keepaspectratio,trim=20  10  10  10,clip]{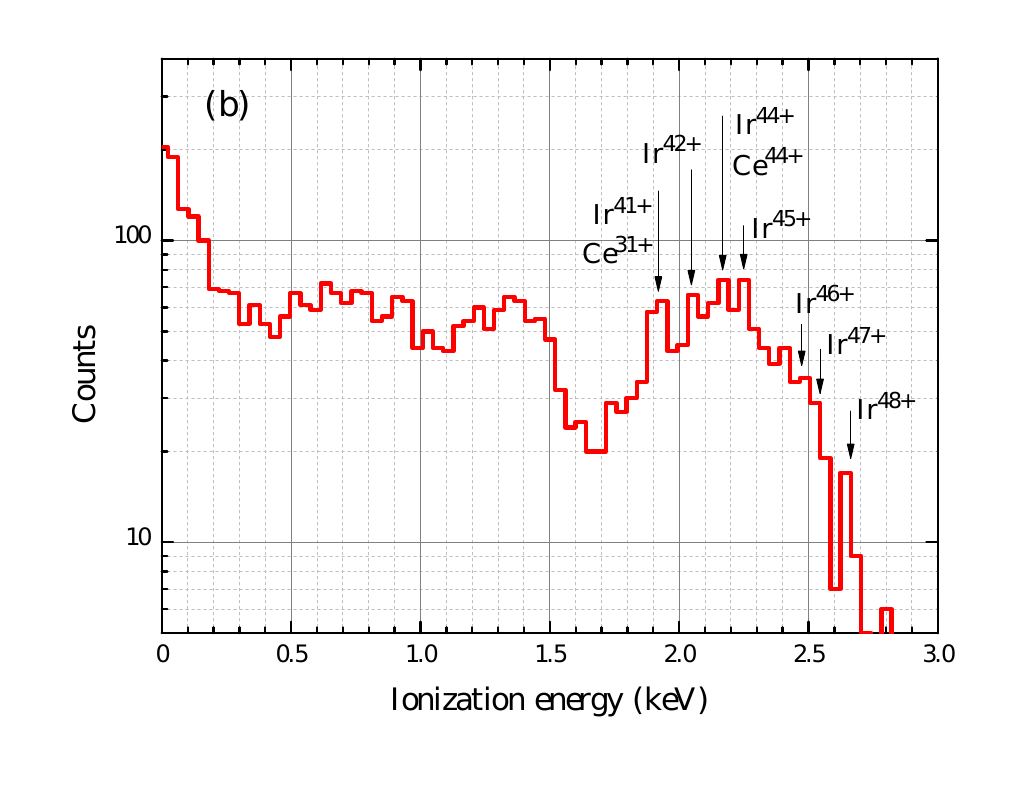}
\end{minipage}
\caption{\label{fig10} (Color online) X-ray radiation of iridium and cerium (cathode materials) from the ion trap (a) and spectrum due to the radiative recombination with the beam electrons (b). The ionization energy is defined as difference between the energy of emitted photons and the electron beam energy $E_e = 4$~keV. }
\end{figure}

In  Fig.~\ref{fig10}(a), we present the primary results of measurements of X-ray emission, which were performed for electron beams characterized by the current of 40-50~mA and the electron energy of 4~keV. The basic vacuum was not better than $1.2\times 10^{-8}$~mbar. The resolution of the detector is about 100 eV. In the experiments, we observed the stationary plasma, in which the balances of energies and charges for different components of plasma had been already achieved. This means that the process of ion cooling, when highly charged ions push low charged ions out of the ion trap, has significant influence on the resulting spectra. Identification of ion charge states is carried out by analysis of the X-ray spectrum corresponding to the process of radiative recombination of ions with the beam electrons [see Fig.~\ref{fig10}(b)]. The arrows indicate the theoretical positions of radiative recombination peaks of highly charged ions with respect to the ionization energy. The radiation of iridium ions with the charges of up to $+48$ can be identified.

\begin{figure}[tbhp]
\begin{minipage}[t]{\textwidth}
\centering\includegraphics[width=0.65\columnwidth,angle=0,keepaspectratio,trim=20  10  10  10,clip]{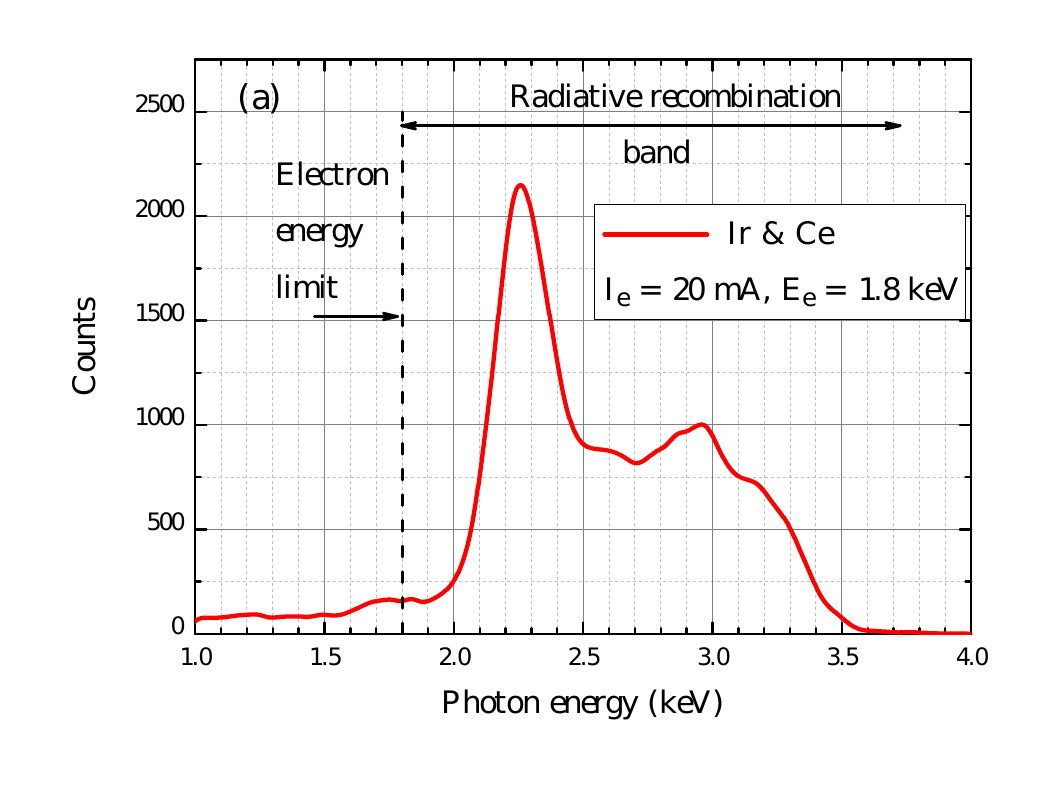}
\end{minipage}
\begin{minipage}[b]{\textwidth}
\centering\includegraphics[width=0.65\columnwidth,angle=0,keepaspectratio,trim=20  10  10  10,clip]{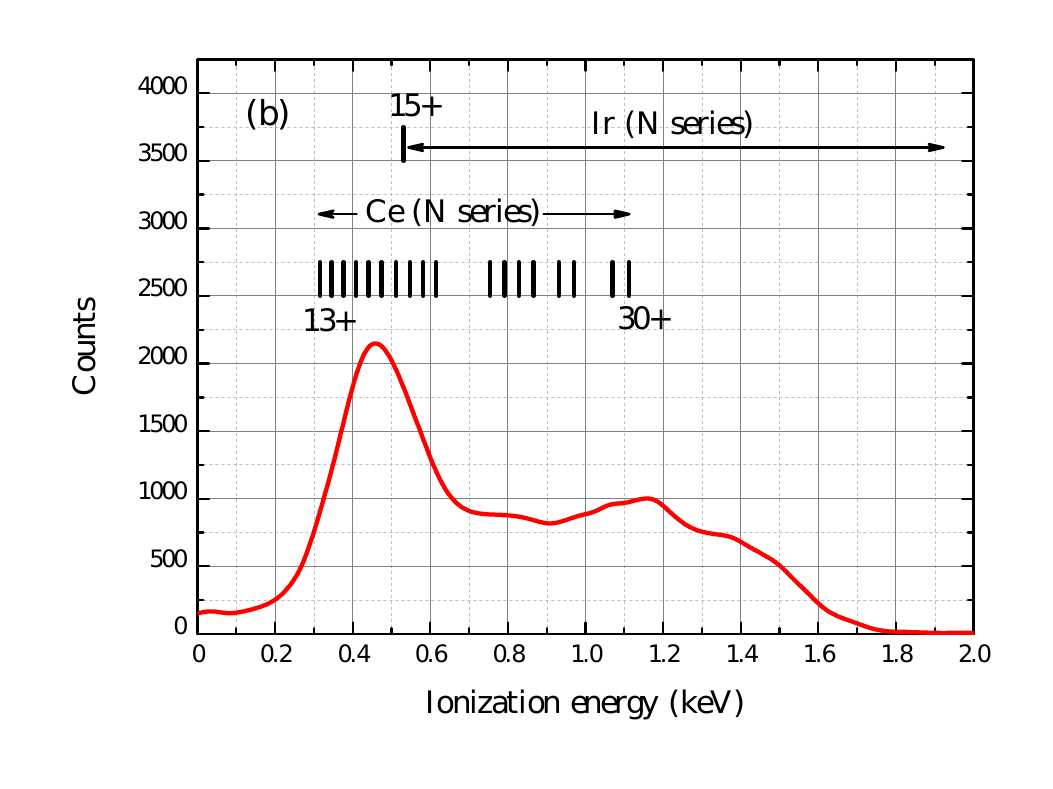}
\end{minipage}
\caption{\label{fig11} (Color online) Similar to Fig.~\ref{fig10} for electron beam with the energy $E_e$ of 1.8 keV.}
\end{figure}

\subsection{Electron beam with the energy of 1.8 keV}

During recent years, a few compact electron beam ion traps (EBITs) operating in the low-energy regime have been constructed for spectroscopic studies of low and moderate  charge state ions  \cite{7,8,9,10}. Although these EBITs are called compact, this is the case in comparison with the EBITs with supercoducting magnet system only. The hand-size device presented in this work is really tiny, being the smallest ion source in the world.

The X-ray spectrum presented in Fig.~\ref{fig11}  is preliminary, because the measurements are made for the same electron-optical system as in the case of electron beam with the energy of 4 keV. In addition, the detector available for the experiment is characterized by lower threshold of registration of about 2 keV, which is rather high. Nevertheless, the spectra demonstrate the ability of this source to apply for spectroscopy of low and moderate charge state ions.  One can argue that Ce ions with charges within the range from $+13$ up to $+30$ and Ir ions with charges larger than $+15$ are produced in the ion trap. The setting can be further optimized for more efficient registration of radiation in the low-energy range.

\section{Injection of Ar into the ion trap}

\begin{figure}[tbp]
\centering
\includegraphics[width=0.65\columnwidth,angle=0,keepaspectratio,trim=20  10  10  10,clip]{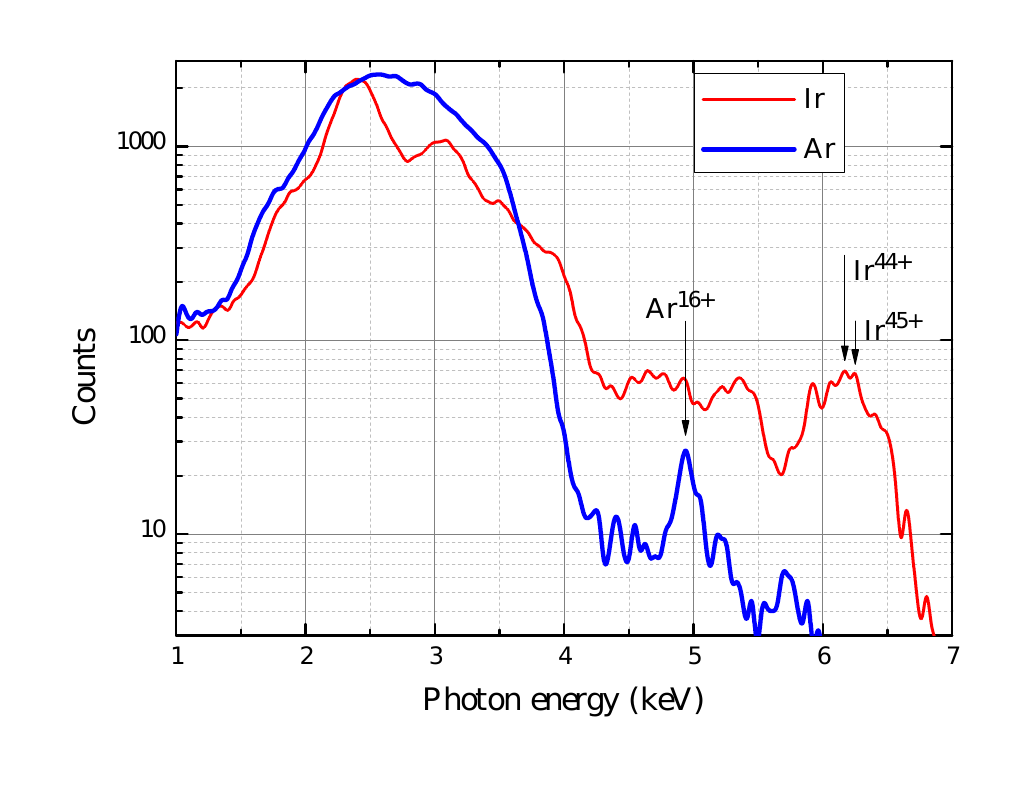}
\caption{\label{fig12} (Color online) X-ray radiation spectrum of Ar ions compared to the basic emission by ions of cathode materials.}
\end{figure}

The equilibrium plasma is produced in the ion trap so long as the electron beam is emitted. In the case of the direct-current (DC) electron beam, the observation of X-ray emission from the installation has some particularities for the elements with atomic numbers lower than that of the cathode material. The problem consists in the following. For plasma being in the thermodynamic equilibrium, there is the limiting concentration of ions of the cathode materials (in our case, iridium and cerium), which is stored in the local ion trap due to significant cathode sputtering under the ion bombardment. Highly charged ions of the cathode material push low charged ions of the working gas out of the trap due the process of ``evaporative cooling''  \cite{11,12}.  If quality of the vacuum is poor enough, the concentrations of Ir and Ce ions in the trap become significantly greater than that of light ions in low charged states. It implies that the ion source should run under high vacuum. In particular, the production of Ar ions in the ion trap becomes efficient only, if the basic vacuum is made better than $1\times 10^{-8}$~mbar. A comparison of the general X-ray spectrum of argon ions with the basic radiation spectrum of iridium and cerium ions is presented in Fig.~\ref{fig12}.

\begin{figure}[tbp]
\centering
\includegraphics[width=0.65\columnwidth,angle=0,keepaspectratio,trim=20  10  10  10,clip]{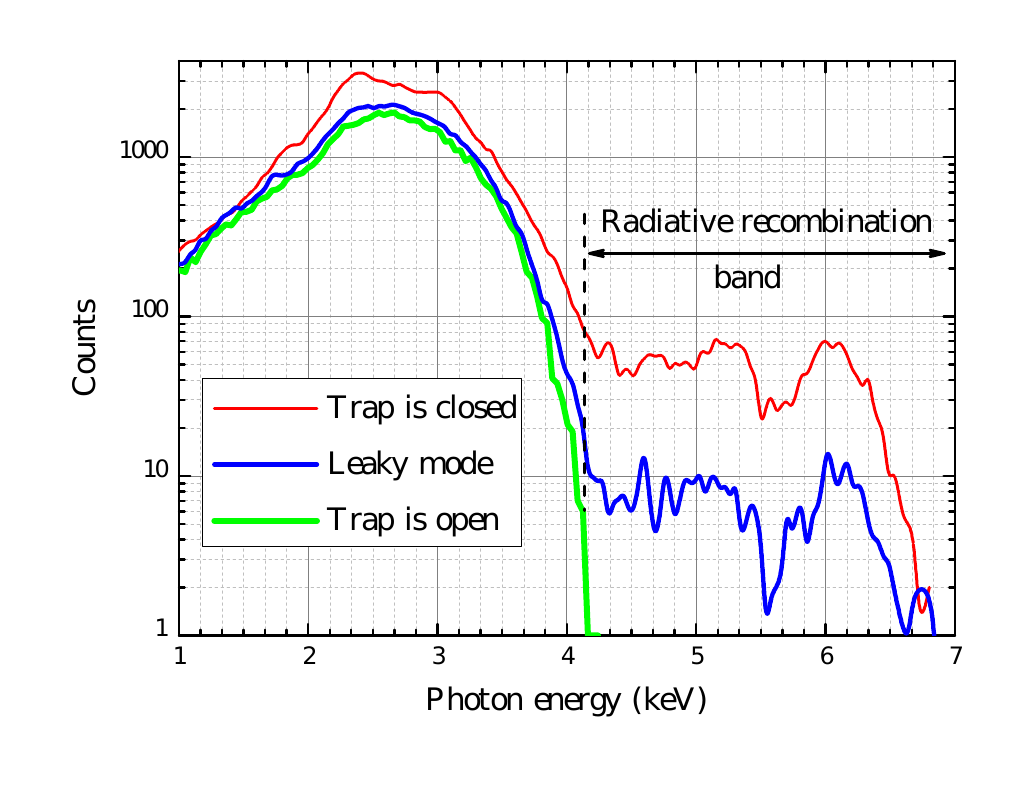}
\caption{\label{fig13} (Color online) X-ray spectrum of Ir and Ce ions evidencing the radial extraction.}
\end{figure}

\section{Ion extraction in the perpendicular direction}

The radial extraction of ions was tested by analysis of the X-ray emission due to radiative recombination of ions of the cathode materials with the beam electrons. The experimental results are shown in Fig.~\ref{fig13}. As follows from Fig.~\ref{fig2}(a), the trapping regime is realized in the case of equality of potentials of the extractor electrode and the drift tube (anode). The appearance of ion trap is confirmed by the presence of the X-ray radiation within the radiative recombination band (see solid red line in Fig.~\ref{fig13}). When potential of the extractor becomes negative with respect to potential of the anode, the radiation intensity decreases. In this case, the leaky mode takes place  (see solid blue line in Fig.~\ref{fig13}). Increasing the negative voltage on the extractor leads to complete disappearance of the recombination radiation from the trap. This means that the ions do not store in the ion trap anymore, but leave it. Thus the possibility of ion extraction from the local ion trap in the radial direction can be considered proven technology.

\begin{figure}[tbhp]
\includegraphics[width=0.4\columnwidth,angle=0,keepaspectratio,trim=0  -100  0  330,clip]{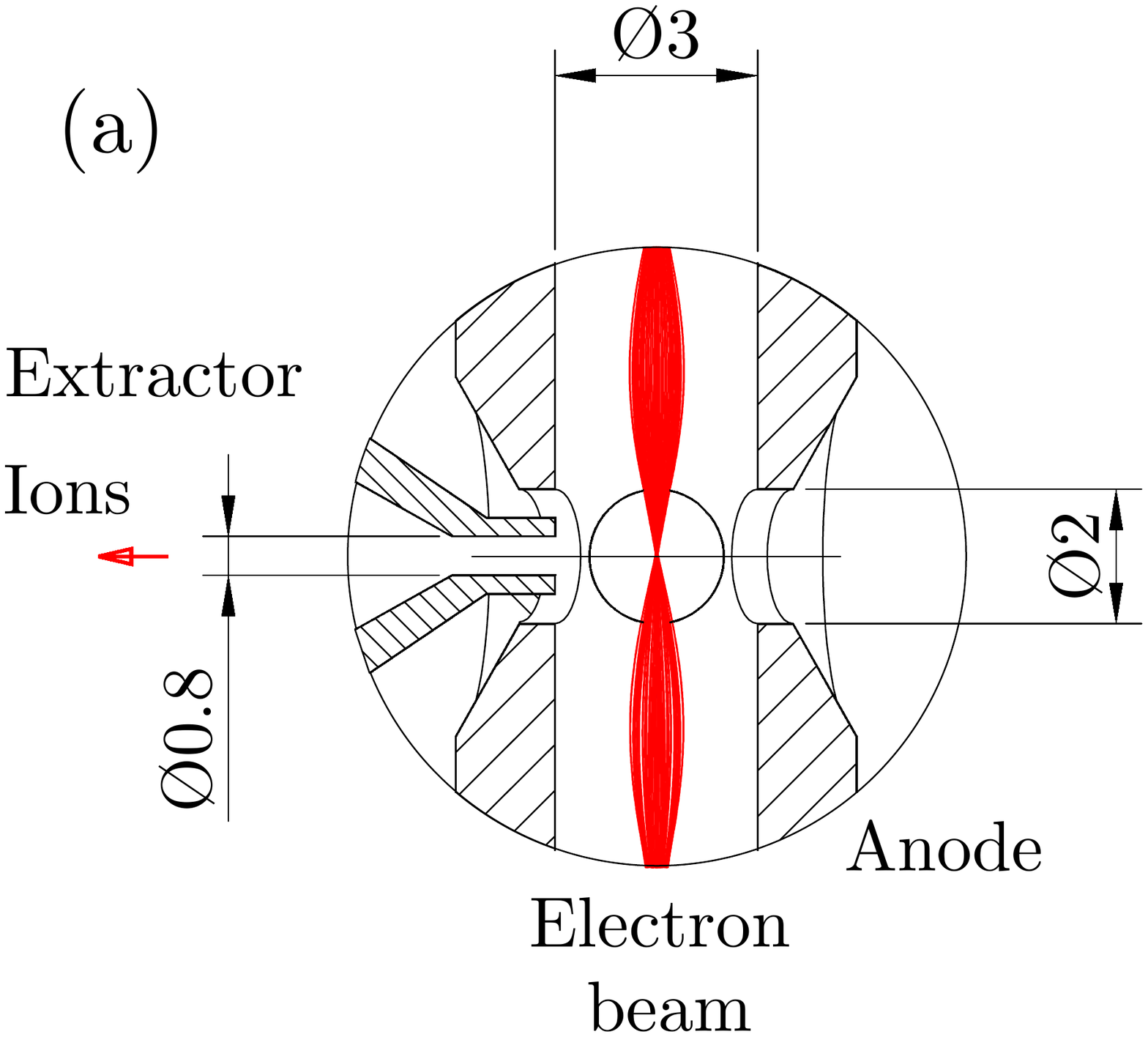}
\hfill
\includegraphics[width=0.57\columnwidth,angle=0,keepaspectratio,trim=20  10  10  10,clip]{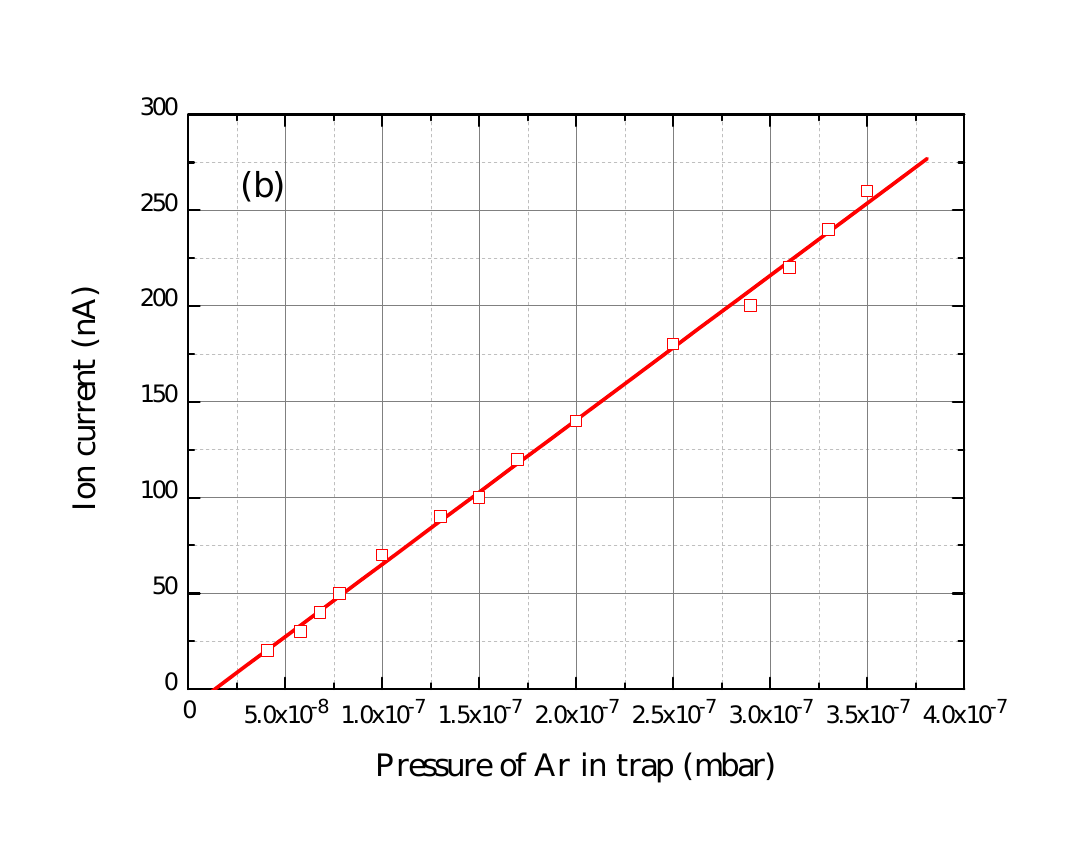}
\caption{\label{fig14} (Color online)  Geometry of DC voltage extraction region (diameters are given in mm) (a) and ion yield as  function of pressure of Ar in the installation (b).}
\end{figure}

\section{DC voltage extraction running mode}

The running mode with DC voltage extractor allows one to obtain relatively high brightness of gaseous ions due to extremely small volume occupied by plasma. In the preliminary experiments, we investigated ability of the ion source to operate in a relatively low vacuum and estimated the total ion current. The ion collector was installed at a distance of approximately 2 cm from the center of the ion trap. The geometry of extraction region is shown in Fig.~\ref{fig14}(a). The input diameter of extractor is equal to  0.8 mm, while the distance from the input of extractor to the center of ion trap is about 1.5 mm. The argon is taken as working gas. The ex\-pe\-ri\-men\-tal dependence of ion current on the quality of vacuum in the installation for the electron current $I_e$ of 30 mA and the electron energy $E_e$ of 4 keV is shown in Fig.~\ref{fig14}(b).

\section{Conclusions}

The novel room-temperature ion source of the next generation operating in different regimes is designed for a wide field of applications. The device is extremely compact, reliable, convenient and cheap in maintenance. In spectroscopic studies, the tiny length of the ion trap compared to the superconducting EBIT at least does not make the efficiency of X-ray detection worse, because the distance between the ion trap and detector is also reduced significantly. In the extraction running mode, the smallness of volume of the ion trap can be more than compensated by high repetition rate due to huge current density of the electron beam. Finally, we note that position of the main (most acute) focus of the rippled electron beam exhibits strong dependence on the electron energy and the characteristics of focusing magnetic field. For particular distribution of the magnetic field, there are only discrete values of energy of the electron beam, which are suitable for operation of the MaMFIS in the ion extraction mode. This problem is solved by more intelligent design with the adjustable cathode and the anode position without breaking vacuum and modular design of the focusing magnetic system.

\section*{Acknowledgements}

The authors express deep gratitude to A.~M\"{u}ller for giving opportunity to test the MaMFIS in the laboratory of Institute of Atomic and Molecular Physics (Justus-Liebig University of Giessen) and to A. Borovik Jr., K. Huber and H.-J. Sch\"{a}fer for their support in X-ray measurements.

\end{document}